\documentclass[graybox]{svmult}

\usepackage{type1cm}        
\usepackage{makeidx}         
\usepackage{graphicx}        
\usepackage{multicol}        
\usepackage[bottom]{footmisc}
\usepackage{newtxtext}       %
\usepackage{newtxmath}       
\usepackage{amsmath}
\makeindex             

\def\ptl{\partial}
\def\ie{\textit{i.e.}, }
\def\eg{\textit{e.g.}, }
\def\kB{k_\mathrm{B}}
\def\glv{\gamma_\mathrm{LV}}
\def\gsl{\gamma_\mathrm{SL}}
\def\gsv{\gamma_\mathrm{SV}}
\def\NA{N_\mathrm{A}}
\def\ds{\displaystyle}
\def\mrv{\mathrm{V}}
\def\mrl{\mathrm{L}}
\def\mrlv{\mathrm{LV}}
\def\mrb{\mathrm{bulk}}
\def\eqref#1{(\ref{#1})}
\def\angb#1{\left<#1\right>}
\def\bm#1{\mbox{\boldmath $#1$}}
\def\rhoL{\rho_\mathrm{L}}
\def\rhoV{\rho_\mathrm{V}}
\begin{document}

\title*{Introduction to Molecular-Scale Understanding of Surface Tension}
\author{Yasutaka Yamaguchi, Hiroshi Kawamura}
\institute{Yasutaka Yamaguchi \at Department of Mechanical Engineering, 2-1 Yamadaoka, Suita 565-0871, Japan
\email{yamaguchi@mech.eng.osaka-u.ac.jp}
}
%
%
\maketitle
\abstract{
In this article, microscopic understanding of the surface tension are provided, which needs basic knowledge of thermodynamics, statistical mechanics as well as continuum mechanics.  
By introducing the intermolecular interaction potential and temperature definition, and by showing conceptual pictures including some results obtained by molecular dynamics simulations, the author hopes that the target readers of undergraduate level students can find fascinating aspects of surface tension as the boundary of macroscopic and microscopic physics.
}
\section{Introduction}
\label{sec:1}
\begin{figure}[b]
\sidecaption
\includegraphics[scale=.80]{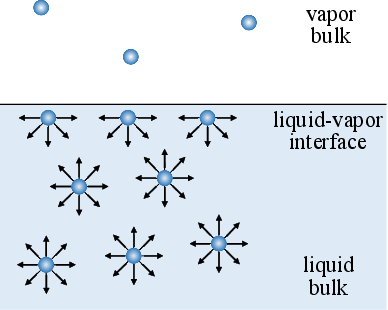}
\caption{Schematic for the molecular scale explanation 
of the surface tension.}
\label{fig:interface}       
\end{figure}
Liquids are supposed to be imcompressible due to their small 
compressibility, \eg its value of water is about 
$5\times 10^{-10}$~Pa$^{-1}$.  Thus, when liquids are in an 
open space or in a closed space with a volume larger than 
theirs, they dispose solid-liquid and/or solid-gas interfaces, 
hence, liquids we see in our daily life almost always have 
interface. Interfacial tensions are termed as the force 
exerted on such interfaces.
\par
At present, the existence of molecules is commonly 
accepted, and the mechanism of the liquid-vapor (LV) or 
liquid-gas (LG) surface tension, known as a tensile force 
at the interface, is 
usually explained by using a schematic as exemplified 
in Fig.~\ref{fig:interface}.
As in this figure, the molecules around the liquid-vapor 
interface have less partners to interact with than those in liquid 
bulk, and the 
liquid-vapor interface is generally disadvantageous to liquid bulk.
%
This seems to be simple and intuitively understandable; however, 
as Ref.~\cite{Marchand2011} pointed out, Fig.~\ref{fig:interface} 
with the arrow directing outward the molecules gives an impression 
that molecules in the liquid bulk and at the liquid-vapor interface 
`pull' each other, and the molecule at the interface is subject 
to a net force along the direction perpendicular to the interface, 
not a force parallel to the interface. This kind of confusing 
questions are often posed especially for interfacial phenomena, 
and indeed, I personally think this is one of the fascinating 
aspects of interface physics which includes macroscopic 
and microscopic features.
\par
In this article, basic pictures for understanding of the surface 
tension are provided with some simulation results, 
This concept is similar to  Refs.~\cite{Marchand2011} and 
\cite{deGennes2003}, basically targeting undergraduate level
students, contrary to many other textbooks on this subject; 
however, I believe that basic but wide and throughout outlook
of thermodynamics, statistical mechanics and continuum mechanics 
is needed to understand the physics of interface, and 
more importantly, to avoid misleading which also have trapped 
professional scientists~\cite{Gao2009}.
Partly related to this point, molecular 
dynamics (MD) simulations have become common especially 
with the development of high-performance computers and 
simulation packages including LAMMPS~\cite{lammps} and 
GROMACS~\cite{gromacs} from the beginning of the 21st 
century, and we can easily have access to the movies of 
various (colorful) molecules. The great scientists in 
the history could not have such movies, and they had 
to imagine basically from macroscopic experimental results; 
however, this does not mean that we will not be trapped 
by the misunderstanding. To advance science, we have 
to make use of both the robust theories constructed 
by the great scientists in the history and computer 
simulations to extend our understanding.
\section{Brief history}
\label{sec:history}
\begin{figure}[b]
\sidecaption
\includegraphics[scale=.80]{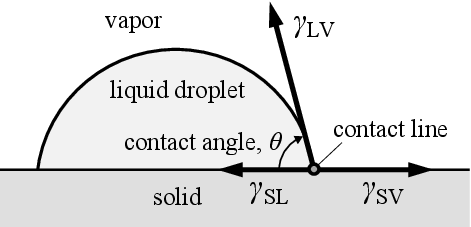}
\caption{Schematic of Young's equation proposed as the surface lateral 
force balance of interfacial tensions exerted on the contact line.}
\label{fig:Young}       
\end{figure}
Although there are several opinions about the origin of the study of capillarity or surface tension~\cite{deGennes2003,Rowlinson2002,Ono1980}, scientific and mathematical approach toward its understanding through the modeling of wetting started from the beginning of the 19th century with T. Young (1773--1829) and P.-S. Laplace (1749--1827). In 1805, Young proposed the following equation~\cite{Young1805}
\begin{equation}
\gsl - \gsv +\glv \cos \theta = 0,
\label{eq:Young}
\end{equation}
where $\gsl$, $\gsv$ and $\glv$ are the solid-liquid, solid-vapor, and liquid-vapor interfacial tensions, repectively, and $\theta$ is the contact angle as the 
angle between the SL and LV interfaces. 
Equation~\eqref{eq:Young} is called Young's equation. 
Although his original article~\cite{Young1805}, indeed in ``essay" style, has only one table and gives no equation nor figure, Eq.~\eqref{eq:Young} was originally suggested to express the force balance 
parallel to the solid surface exerted on the contact 
line as shown in Fig.~\ref{fig:Young}. 
Note that the proposal of Young's equation~\eqref{eq:Young} is sometimes referred to as in 1804, because he made a talk about this topic in this year.
Young's equation~\eqref{eq:Young} was extended to evaluate wettability of a 
liquid on a solid surface, known as the Young-Dupr\'{e} equation:
\begin{equation}
    S = \glv (\cos \theta - 1),
    \label{eq:spr_cos}
\end{equation}
where the spreading coefficient $S$ given by
\begin{equation}
    S = \gsv - (\gsl + \glv)
    \label{eq:spreadingcoeff}
\end{equation}
expresses the wettability: $S=0$ gives $\cos \theta=1$ in 
Eq.~\eqref{eq:spr_cos} and positive $S$ means that the liquid completely covers the solid surface.
%
%
\par
The year of 1804 or 1805 was before the establishment of thermodynamics: N. L. S. Carnot (1796--1832) was under ten, and J. P. Joule (1818--1881), W. Thomson (Lord Kelvin, 1824--1907)
R. Clausius (1822--1888), and  J. W. Gibbs (1839--1903) 
were not born, and this clearly means that Young did not and could not bring the concept of thermodynamics nor molecular interaction into his modelling of wetting and surface tension as shown in Fig.~\ref{fig:interface}~\cite{Gao2009}. 
Note that the primary hydrodynamic description summarized 
in ``Hydrodynamica'' by D. Bernoulli (1700--1782) \cite{DBernoulli1738},
which was based on Newtonian mechanics by I. Newton (1642--1727), 
was already published at that time, and the Euler 
Equations about ideal fluids by L. Euler (1707--1783)
as the mathematical framework of fluid mechanics
as well as the Lagrangian mechanics by J.-L. Lagrange (1736--1813)
were already available~\cite{Fujiwara1995}. 
Probably, continuum mechanics including the concept of 
stress tensor was not available considering the 
ages of A.-L. Cauchy (1789--1857), C. L. M. H. Navier 
(1785--1836) and S. D. Poisson (1781--1840).
\par
Anyway, by the frontiers including the above-mentioned scientists, thermodynamics was established in the 19th century, and based on it, the van der Waals equation of state expressing the phase coexistence was proposed by J. D. van der Waals (1837--1923). He also wrote an article about surface tension in 1893~\cite{vdW1893,Rowlinson1979}, and  introduced the concept of dividing surface and formulated a thermodynamic framework with a liquid-vapor interface.
Just after the tragedy of L. Boltzmann (1844--1906), J. Perrin (1870--1942) experimentally proved the existence of molecules inspired by the idea of A. Einstein (1879--1955)~\cite{Fujiwara1995}, and this enabled the above-mentioned explanation of the surface tension through the interaction potential between molecules 
interacting with a potential well depth, \ie with a moderate attraction force for long intermolecular distance range and a strong repulsion for short distance range described in detail below.
Based on this interaction potential model,  G. Bakker (1856--1938), R. C. Tolman (1881--1948), 
J. G. Kirkwood (1907--1959),  F. P. Buff (1924--2009), etc. constructed the theoretical framework 
of the surface tension in thermodynamic equilibrium.
Note that Japanese scientists including S. Ono (1918--1995), S. Kondo (1922--2014), A. Harashima (1908--1986), 
etc. largely contributed to the development. Especially, Ono and Kondo wrote a great review~\cite{Ono1960}, 
and they categorized the approaches of surface tension into thermodynamic and quasi-thermodynamic ones, statistical mechanical ones and hydrodynamic 
(also mentioned as ``mechanical'' in \cite{Ono1980}) ones. 
Ono also left an instructive textbook in Japanese~\cite{Ono1980}, and he wrote 
in the introduction of the book: 
``surface is very thin but still has thickness, and it is a pity that its physics is difficult to understand because of this fact.'' This sentence  clearly points out the difficulty of surface, and also shows the fascinating feature of surface physics.
Another great textbook by J. S. Rowlinson and B. Widom~\cite{Rowlinson1982} is also available; indeed, the history in the present article mentioned above is based on these books~\cite{Ono1980,Rowlinson1982}.
\par
As described above, it is basically necessary to include the microscopic concept to fully explain or understand the mechanism of surface tension and wetting. On the other hand, as we see water droplets on solid surfaces almost everyday, such a liquid motion affected by the surface tension is very common and  macroscopic, \ie visible by bare eyes. As Young modelled, it is possible to understand and predict the macroscopic liquid behavior by simply considering the surface tension as the force to reduce the surface area without the above-mentioned microscopic knowledge.
Indeed in the middle of 18th century, an experimental approach to measure the surface tension was proposed by L. F. Wilhelmy (1812--1864).
In this method called the Wilhelmy plate method~\cite{Wilhelmy1863}, which is commonly used 
also at present, a solid plate or a cylinder is vertically immersed into a liquid pool, and the surface tension is evaluated by measuring the contact angle and the vertical force exerted on the solid. 
This standpoint is equivalent to the above-mentioned hydrodynamics approach that Ono and Kondo categorized, in which the microscopic physics of interface with non-zero thickness is integrated into the surface tension on a surface with zero thickness. 
As the name indicates, this hydrodynamic implementation matches well with the governing equations of fluid dynamics, and is especially familiar with the numerical simulation technique called the computational fluid dynamics (CFD). In practice, thermodynamic and statistical 
mechanical approaches cannot be applied for dynamic systems in principle because 
these approaches assume static thermodynamic equilibrium.
Anyway, CFD simulations of liquid flow with moving interface are common 
at present owing to the development of high-speed computers from 1980s. 
\par
\begin{figure}[b]
	\sidecaption
	\includegraphics[width=7.5cm]{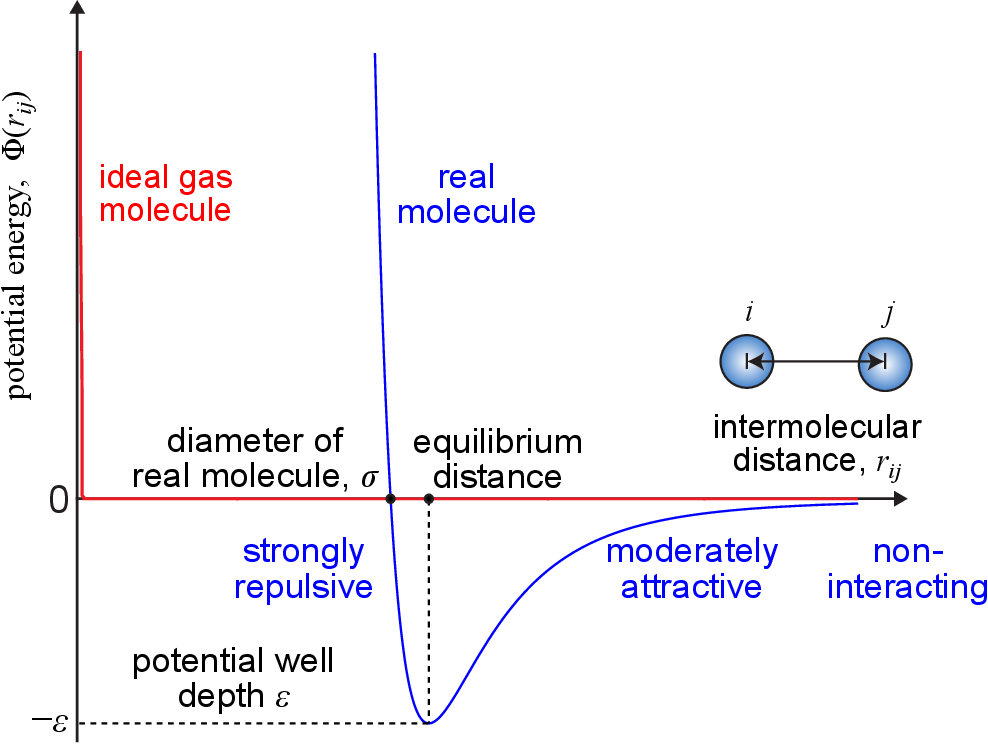}
	\caption{
Intermolecular interaction between ideal gas molecules and 
between real molecules. No intermolecular force acts between ideal 
gas molecules and they have no volume. On the other hand, real 
molecules are generally attracted by each other for an intermediate 
interatomic distance and repelled for a smaller distance around the
molecular diameter, and they have almost no interaction for a 
longer distance. The minimum point of the potential energy is the 
equilibrium distance.
\label{fig:potential}       
}
\end{figure}

Parallel to the CFD, the (classical) molecular 
dynamics (MD) method, which solves the motion of constituent molecules of fluids 
governed by inter-molecular potential functions based on the Newtonian mechanics, 
has also been developed, and this method enabled the simulations 
of liquid behavior with interfaces as a microscopic approach.
Indeed, one can obtain an equilibrium liquid-vapor coexistence 
system with an interface by locating a certain number of 
molecules in a simulation cell, whose intermolecular force 
is given by a potential function exemplified in Fig.~\ref{fig:potential}, 
and by controlling the temperature of the system. 
As pointed out in Ref.~\cite{Rowlinson1982}, this solves 
the main difficulty in the statistical mechanical 
approach of how to analytically obtain the equilibrium density 
distribution in such liquid-vapor coexistence system from the microscopic 
intermolecular potential function.
\section{Why phase separation happens and why interface is formed
-- from thermodynamics --}
%
\label{sec:2}
\subsection{Ideal gas model and temperature}
In order to understand the properties of the liquid-vapor 
interface, a primitive question of why the constituent 
molecules try to gather to form liquid and vapor phases 
having different densities as illustrated in 
Fig.~\ref{fig:interface} 
instead of forming a single homogeneous phase with 
a uniform density. 
Indeed, a fluid which always stays 
single homogeneous phase exists
as a conceptual one: an ideal gas. 
The principal (macroscipic) definition of ideal gases 
is that they always obeys the equation of state:
\begin{equation}
pV = nRT,
\label{eq:ideal_eos}
\end{equation}
%
where $p$, $V$, $n$, $T$ and $R$ denote the pressure, the volume, the amount of substance 
of the gas in moles, the absolute temperature, 
and the universal gas constant, respectively. 
Indeed, Eq.~\eqref{eq:ideal_eos} holds under 
high temperature and low pressure conditions 
for real gases, and an ideal gas model is an 
extended imaginary substance which obeys 
Eq.~\eqref{eq:ideal_eos} irrespective 
of the temperature and pressure.
D. Bernoulli was the first to propose the basis for the 
kinetic theory of gases  in his book of Hydrodynamica~\cite{DBernoulli1738}
in the 18th century. In this book, he argued 
that a gas consists of a huge number of molecules moving in all directions, 
and their impact on a surface causes the 
pressure of the gas, and thus their average kinetic energy 
determines the temperature of the gas under the following assumptions:
\begin{itemize}
\item
The constituent molecules of the gas are infinitesimally small hard spheres, besides they are subject to elastic collisions among each other and with the surroundings (container wall).
\item
There are no attractive or repulsive forces between the molecules apart from those that determine their point-like collisions, \ie 
the only forces between the gas molecules and the surroundings are impulsive force upon the point-like collisions.
\item
The molecules are constantly moving, and as a result of 
multiple random collisions, their velocity distribution 
reaches a certain isotropic random directions as an 
equilibrium state.
\end{itemize}
Almost after a century, the idea was revisited in the 19th 
century by Clausius and J. C. Maxwell (1831--1879), etc.
From the assumption above, it is possible to relate 
the mean square velocity of the constituent 
molecules $\left<|\bm{v}|^{2}\right>$ 
with the pressure 
$p$ with the density $\rho$ of the gas
in three-dimension as
\begin{equation}
    p = \frac{\rho \left<|\bm{v}|^{2}\right>}{3}
    \label{eq:p_rho_msv}
\end{equation}
by considering the sum of impulse exerted from the molecules 
on the container wall of unit area per unit time, where  
$\left<|\bm{v}|^{2}\right>$ 
is defined as the space, temporal and molecular average by
\begin{equation}
\left<|\bm{v}|^{2}\right>\equiv
    \lim_{t_{\infty}\rightarrow \infty} \frac{1}{t_{\infty}}
    \int_{0}^{t_{\infty}} 
    \left[
\frac{1}{N}\sum_{i=1}^{N}|\bm{v}_{i}|^{2}
\right]dt
\end{equation}
for $N$ molecules with the velocity of 
$i$-th molecule being $\bm{v}_{i}$. Note that
$\bm{v}_{i}$ satisfies
\begin{equation}
    \frac{1}{N}\sum_{i=1}^{N}\bm{v}_{i} = \bm{0}
    \label{eq:velave=0}
\end{equation}
for present equilibrium gases, \eg confined in a 
closed container. In case the container is moving 
at a constant velocity, $\left<|\bm{v}|^{2}\right>$ must 
be defined by using the molecular velocities relative 
to the group motion. 
By the way, the relation in Eq.~\eqref{eq:p_rho_msv} 
derived by Clausius is amazing because 
the speed of invisible molecules can be estimated 
only by the two measurable macroscopic values of the 
pressure $p$ and the density $\rho$~\cite{Fujiwara1995}.
%
By inserting Eq.~\eqref{eq:p_rho_msv} into
Eq.~\eqref{eq:ideal_eos}, it follows
\begin{equation}
    M \left<|\bm{v}|^{2}\right> = 3RT,
    \label{eq:mvsq=3RT}
\end{equation}
where $M$ denotes the mass of gas per mole. Then, 
let $m$ and $\kB$ be the mass of single molecule 
and the Boltzmann constant, respectively defined by
\begin{equation}
    m \equiv \frac{M}{\NA}, \quad
    \kB \equiv \frac{R}{\NA} = 1.38\times10^{-23}\ {\rm J/K}
\end{equation}
using the Avogadro number $\NA$, we obtain a 
fundamental relation between the kinetic energy 
of the constituent molecule and the absolute 
temperature:
\begin{equation}
    \frac{1}{2}m\left<|\bm{v}|^{2}\right> 
    =
    \frac{3}{2}\kB T.
    \label{eq:defT_micro}
\end{equation}
Equation~\eqref{eq:defT_micro} can be extended
to real molecules as the microscopic definition 
of temperature $T$.
\subsection{Intermolecular potential}
The intermolecular interaction between ideal gas molecules
can be schematically illustrated in Fig.~\ref{fig:potential}
with the red line. The horizontal line for overall distance 
except at zero distance shown with the vertical line
indicates that the molecules have no interaction force 
except the hard point-like collision. 
In contrast to this ideal gas feature, the 
interaction potential 
between real molecules can be modeled as the blue line
in Fig.~\ref{fig:potential}. Two real molecules within 
an intermediate intermolecular distance attract
each other whereas a strong repulsive force acts
between the two within a smaller distance around the
molecular diameter, and they have almost no interaction 
for a longer distance. The distance giving the minimum 
potential energy of $-\varepsilon$ is the equilibrium 
distance.
\par
\begin{figure}[b]
\sidecaption
\includegraphics[width=100mm]{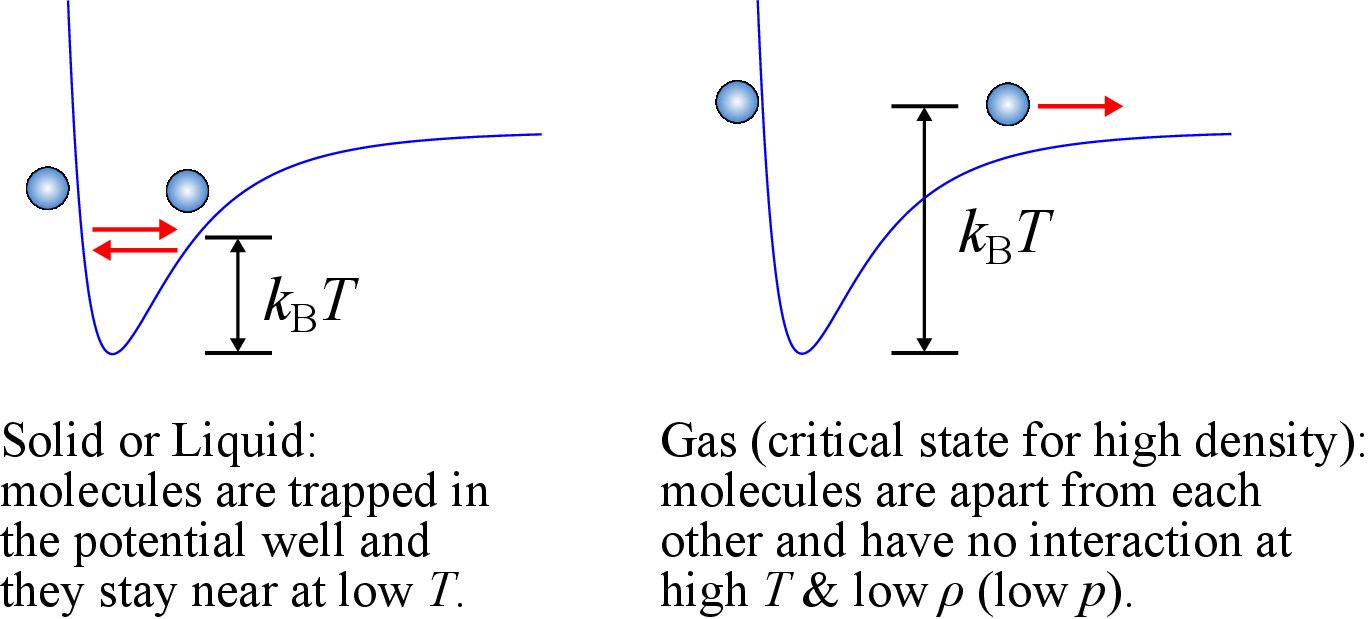}
\caption{Schematic of the intuitive understanding of the 
temperature-dependent phase change from intermolecular 
interaction between real molecules. Left: molecules are 
trapped in the potential well because the average kinetic
energy corresponding to $\kB T$ is not sufficiently large.
Right: molecules can escape from the potential well with
the kinetic energy and freely moving in a ballistic 
manner. The left 
corresponds to the solid or liquid phase whereas the right 
is the gas phase.
}
\label{fig:phasechange}       
\end{figure}
Now we try to intuitively understand phase change from 
this interaction potential of real molecules with a potential 
well depth and the definition of absolute temperature $T$ 
in Eq.~\eqref{eq:defT_micro}. Figure~\ref{fig:phasechange} 
shows the schematic for the understanding of the 
temperature-dependent phase change from the intermolecular
interaction between real molecules. Suppose two molecules 
vibrating around the potential well at a low temperature
without total translational motion as in the left panel, 
\ie satisfying Eq.~\eqref{eq:velave=0}.
Then, as indicated by Eq.~\eqref{eq:defT_micro}, the two molecules 
have certain relative average kinetic energy with 
a scale of $\kB T$. 
\par
At a low temperature with the corresponding 
kinetic energy 
sufficiently 
smaller than the potential well depth $\varepsilon$, the 
molecules are trapped in the potential well. In that case,
molecules rarely change their interaction pairs and 
if there more than two molecules, they try to maximize 
the number of neighboring pairs to minimize the potential 
energy, and they eventually form a solid crystal typically 
with a closed packing structure. 
Note that thermal expansion of solid can also be understood 
from the asymmetric vibration feature around the 
equilibrium point; the intermolecular distance at $T=0$ 
is apparently at the minimum of the potential well while 
the time-averaged mean distance at $T>0$ becomes 
longer as the temperature increases.
This expansion is not expected if the intermolecular 
interaction is modeled by a simple harmonic spring 
connecting the two molecules, which gives a symmetric 
vibration around the point of minimum potential energy.
\par
Under a moderate temperature, molecules can frequently 
escape from the potential well and change the pairs 
without having certain fixed structures. This can be 
understood as liquid. 
At high temperature as shown in the right panel of 
Fig.~\ref{fig:phasechange},
molecules are not trapped in the potential well due 
to their sufficient kinetic energy and freely moving 
in space. This corresponds to the gas phase. 
\par
Hence, the key features of real molecules 
can be summarized as follows:
\begin{itemize}
\item
There are short-range repulsive and 
long range attractive forces between the 
 molecules to form a potential well.
%
\item
The molecules are constantly moving, and as a result of 
multiple random collisions, their velocity distribution 
reaches a certain isotropic random directions as an 
equilibrium state. 
\end{itemize}
The second feature is the same as ideal gases
and the kinetic energy of the random motion is equally 
distributed to each degree of freedom: the equipartition
under thermal equilibrium as a primitive basis of 
statistical mechanics. It is easy to imagine 
why Eq.~\eqref{eq:ideal_eos} holds under high 
temperature and low pressure conditions 
for real gases from the intermolecular 
interaction potential of real molecules; however, 
the history of thermodynamics and statistical 
mechanics tells that constructing it from ideal 
gas potential is not that easy. Indeed, the outstanding 
idea of D. Bernoulli that gases as fluid consist 
of huge number of infinitesimally small molecules, 
proposed in the first half 18th century was not further 
investigated for about a century until Clausius and Maxwell~\cite{Fujiwara1995}.
\par
\begin{figure}[b]
\sidecaption
\includegraphics[scale=.80]{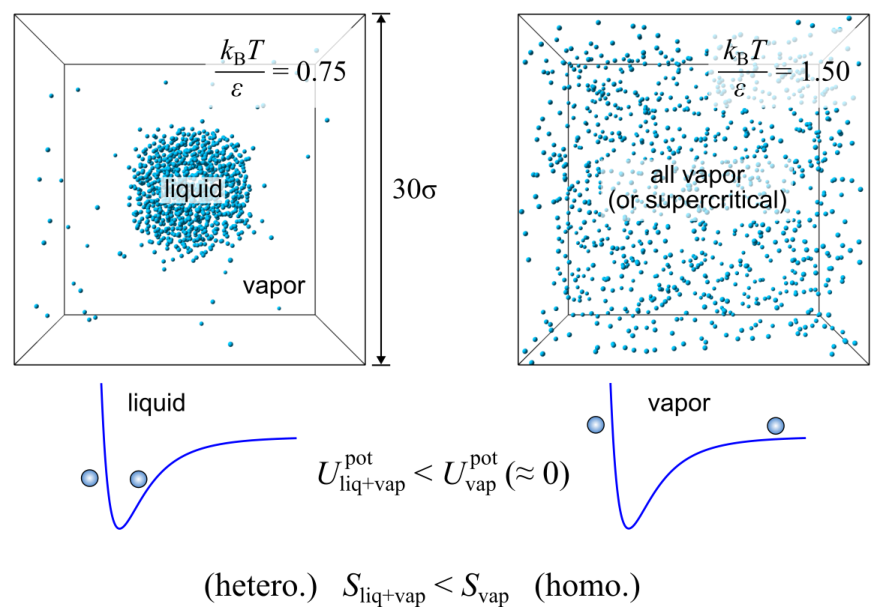}
%
%
\caption{MD simulation results of simple Lennard-Jones molecules confined in a calculation cell.}
\label{fig:drop-homogen}       
\end{figure}
Anyway, now we go back to the interpretation of the 
surface tension in Fig.~\ref{fig:interface}. 
This figure does not mean that the molecular pairs in 
the liquid bulk pull the partner molecule each other 
but they are at a certain mean inter-molecular 
distance as indicated in the right panel of 
Fig.~\ref{fig:potential}.
%
Molecules around the interface have less partners 
to interact with than those in the liquid bulk, 
and those in the vapor bulk have almost no partners.
As mentioned above, with the present 
computers, MD simulations can be easily run even with 
a laptop computers. Figure~\ref{fig:drop-homogen} shows 
the snapshot of equilibrium systems consisting 
of molecules with their intermolecular interaction 
described by the Lennard-Jones potential popularly
used as a simple model expressing the above-mentioned feature:
\begin{equation}
    \Phi_\mathrm{LJ}(r_{ij}) = 
    4\varepsilon\left[
    \left(\frac{\sigma}{r_{ij}}\right)^{12} 
    -
    \left(\frac{\sigma}{r_{ij}}\right)^{6} 
    \right],
    \label{eq:LJ}
\end{equation}
where $\varepsilon$ and $\sigma$ denote the 
potential well depth and diameter shown in 
Fig.~\ref{fig:potential}. One thousand 
molecules are confined in a cubic box 
sized $30\sigma\times 30\sigma \times 30\sigma$ 
with the time-averaged system temperatures, 
\ie kinetic energy given by Eq.~\eqref{eq:defT_micro}, 
set at $\kB T/\varepsilon$ equal to 0.75 and 1.5
in the left and right systems, respectively. 
\par
Note that the molecules in the systems are 
constantly moving as described above, and even 
if one starts a simulation from arbitrary 
initial configuration, for instance, with 
locating the molecules at grid points, the molecules 
spontaneously form corresponding equilibrium
states after a certain time as in the snapshots. 
This term `equilibrium'
in the microscopic scale roughly means that the 
molecules are moving with the positions and 
momenta of the constituent molecules changing 
with time, but their apparent feature as a group 
is unchanged. For instance, most of the constituent 
molecules are gathered to form a spherical liquid 
droplet with remaining molecules flying as vapor around 
the droplet in the left system as an equilibrium
state; the constituent molecules of liquid and 
vapor change but the volume as well as the 
spherical structure of the droplet are unchanged.
Similarly, the positions of momenta of the molecules 
change, but they always keep vapor phase in the 
right system after reaching the equilibrium 
state spontaneously achieved after a certain time
irrespective of the initial condition. 
As described later, this spontaneous 
change of the system is due to the  minimization of 
the Helmholtz free energy of the system.
\par
As clearly observed from the figure, 
liquid-vapor coexistence kept at the lower 
temperature of $\kB T/\varepsilon=0.75$ 
in the left panel is lost 
with the temperature rise, and the system
is filled with vapor (indeed supercritical
phase) at the higher temperature of 
$\kB T/\varepsilon=1.5$. This phase change 
can be intuitively understood from the 
schematic 
in Fig.~\ref{fig:phasechange}.
%
\par
We further think about why the droplet in 
the left panel takes the spherical structure
to minimize the surface area from an energetic
point of view.
Let the potential energy per molecule in the vapor
bulk, at the LV interface, and in the liquid bulk be 
$e_\mrv$, $e_\mrlv$ and $e_\mrl$, 
respectively. Then, they satisfy 
\begin{equation}
   e_\mrl < 
   e_\mrlv \ll 
   e_\mrv \approx 0
   \label{ineq:e_per_mol}
\end{equation}
considering the schematic in Fig.~\ref{fig:interface}
and intermolecular potential in Fig.~\ref{fig:potential},
which indicates that $e_\mrv$, $e_\mrlv$ are basically 
negative whilst $e_\mrl$ is almost zero for molecules 
in the vapor phase without partners to interact with.
The first inequality implies that the number 
of neighbouring molecules is smaller at the 
interface than in the liquid bulk.
Consequently 
\begin{equation}
\quad e_\mrv - e_\mrlv > 0
\label{eq:ev-elv>0}
\end{equation}
holds.
In addition, let the number of corresponding 
molecules and volumes 
being 
$N_\mrl$, 
$N_\mrlv$ and 
$N_\mrv$, 
and 
$V_\mrl$, 
$V_\mrlv$ and 
$V_\mrv$, 
respectively. Then, the number of molecules $N$ and volume $V$ 
of the system write
\begin{equation}
    N = 
    N_\mrl
    +
    N_\mrlv
    +
    N_\mrv
\end{equation}
and
\begin{eqnarray}
V 
= 
V_\mrl
+
V_\mrlv
+ 
V_\mrv,
\label{eq:vol}
\end{eqnarray}
respectively. In the macroscopic scale, we assume $N_\mrlv =0$ and $V_\mrlv =0$ while the LV interface is a region in the molecular scale and molecules indeed exist there. We define the corresponding volumes per molecule as
\begin{equation}
v_\mrl = \frac{V_\mrl}{N_\mrl},\ 
v_\mrlv = \frac{V_\mrlv}{N_\mrlv},\ 
v_\mrv = \frac{V_\mrv}{N_\mrv}.
\end{equation}
Assuming that $v_\mrlv$ is approximately 
the same as $v_\mrl$, it follows for the volume per molecule 
that 
\begin{equation}
   0 < v_\mrl \approx 
   v_\mrlv \ll 
   v_\mrv.
\end{equation}
Then, Eq.~\eqref{eq:vol} rewrites
\begin{equation}
V 
=
N_\mrv v_\mrv + + N_\mrlv v_\mrlv + N_\mrv v_\mrl
\approx
(N_\mrl + N_\mrlv) v_\mrl + N_\mrv v_\mrv.
\label{eq:vol_approx}
\end{equation}
%
On the other hand, the internal energy $U$ of a system is 
separated into the kinetic and potential contributions
$U^\mathrm{kin}$ and $U^\mathrm{pot}$, respectively as
\begin{equation}
U = U^\mathrm{kin} + U^\mathrm{pot}.
\label{eq:U=Ukin+Upot}
\end{equation}
Using Eq.\eqref{eq:defT_micro}, the former is given by
\begin{equation}
\ds U^\mathrm{kin} 
\equiv
\sum_{i=1}^{N} \frac{1}{2}m|\bm{v}_{i}|^{2}
=
\frac{3}{2}N\kB T,
\end{equation}
which is constant
under the constant temperature condition. 
The latter of potential term is formally given by
\begin{eqnarray}
\ds U^\mathrm{pot} 
\equiv 
\sum_{i=1}\sum_{j(>i)}\Phi(r_{ij}).
\end{eqnarray}
This can be approximated using the mean 
potential energy per molecule as 
\begin{eqnarray}
\nonumber
\ds U^\mathrm{pot} 
&\approx &
N_\mrv e_\mrv 
+ 
N_\mrlv e_\mrlv
+ 
N_\mrl e_\mrl
\\
&=&
(N_\mrl + N_\mrlv)e_\mrl + N_\mrlv(e_\mrlv - e_\mrl).
\label{eq:upot_sum}
\end{eqnarray}
From Eqs.~\eqref{eq:vol_approx}-\eqref{eq:upot_sum}
and 
Inequality~\eqref{eq:ev-elv>0}, 
it is shown that with keeping $N_\mrl + N_\mrlv$ constant, 
reducing the internal energy $U$ in Eq.~\eqref{eq:upot_sum} 
is possible by decreasing $N_\mrlv$
without changing 
the volume $V$ in Eq.~\eqref{eq:vol_approx}.
This means that the system internal 
energy can be reduced by decreasing 
the surface area $A_\mrlv$ because 
$N_\mrlv$ is apparently proportional to $A_\mrlv$.
This leads to the formation of the spherical 
droplet in the left panel of Fig.~\ref{fig:drop-homogen}
as an equilibrium shape with the smallest surface area 
in the three-dimensional space.
\subsection{Free energy and entropy}
The explanation above looks reasonable; 
however, another question arises why the system 
prefers to take the `all vapor' state with less 
internal energy $U$ at higher temperatures. 
The trap model in 
Fig.~\eqref{fig:phasechange} merely indicates 
that the molecular pairs easily dissociate 
from each other, but they can have a lower 
internal energy even at a higher temperature 
if they form more intermolecular pairs. 
To answer this question, the free energy 
instead of the internal energy must be 
considered. The second law of thermodynamics 
tells us that the Helmholtz free energy $F$ 
of a system given by
\begin{equation}
F\equiv U - TS
\label{eq:def_helmholtz}
\end{equation}
decreases until the 
system reaches the equilibrium state, \ie
\begin{equation}
    dF \leq 0\ (\mathrm{nonequilibrium}),\quad
    F = F_\mathrm{min}\ (\mathrm{equilibrium})
\end{equation}
holds for $F$ for a system under constant volume 
$V$ and constant temperature $T$ condition. This 
corresponds to a closed container in contact 
with a  a constant-temperature heat bath.
The key point is that the Helmholtz free energy 
in Eq.~\eqref{eq:def_helmholtz} 
includes the entropy of the system $S$.
\par
In classical statistical mechanics, the entropy $S$ is 
related to the ``number of possible equivalent 
microstates corresponding
to a certain macroscopic state" $\Omega$ 
as follows:\footnote{
Boltzmann's tombstone bears the inscription of
`$S = k \log W$'.
}
\begin{equation}
    S = \kB \ln \Omega,
    \label{eq:s=klnw}
\end{equation}
where a microstate in 3-dimensional system of $N$-molecules 
is determined by giving 
3$N$-positions and $3N$-momenta of constituent molecules.
From Eq.~\eqref{eq:s=klnw}, it is possible to show that 
the position contribution to entropy $S$ is the largest 
for a system with homogeneous density.
We will see that in the following example:
\par\noindent
 \hrulefill 
\\
\textbf{(Example)}
Suppose a closed equilibrium system with ideal gas molecules in a box
at a constant temperature, 
and let $N_\mathrm{left}$ and $N_\mathrm{right}$ be the numbers of gas 
molecules in the left-half and right-half of a box, respectively, and denote the case by $[N_\mathrm{left},N_\mathrm{right}]$ ($N=N_\mathrm{left}+N_\mathrm{right}$). Evaluate the number of possible cases and compare the two for the following three cases,  respectively:
\begin{enumerate}
\item
Case of small total number $N_\mathrm{total}=200$: 
$[99, 101]$ vs. 
$[100, 100]$
\item
Case of a larger total number $N_\mathrm{total}=2000$:
$[990, 1010]$ vs. 
$[1000, 1000]$
\item
Case of a huge (Avogadro scale) total number of $N_\mathrm{total}=2\times 10^{23}$:
$[0.99\NA, 1.01\NA]$ 
vs. 
$[N_\mathrm{left}=\NA, N_\mathrm{right}=\NA]$ 
\end{enumerate}
\textbf{(Answer)} The numbers of 
cases $\Omega$ to separate 
$N_\mathrm{left} + N_\mathrm{right}$ 
into left and right are given by
%
\begin{enumerate}
\item 
$N_\mathrm{total}=200$:
\begin{eqnarray*}
& \ds 
\Omega_{[99, 101]}
=
\frac{200!}{99!101!},
\quad
\Omega_{[100, 100]}
=
\frac{200!}{100!100!}
\\ & \ds 
\frac{
\Omega_{[99, 101]}
}{
\Omega_{[100, 100]}
}
= 
\frac{100!100!}{99!101!}
=
\frac{100}{101}
\end{eqnarray*}
\item 
$N_\mathrm{total}=2000$:
\begin{eqnarray*}
& \ds
\Omega_{[990, 1010]}
=
\frac{2000!}{990!1010!},
\quad
\Omega_{[1000, 1000]}
=
\frac{2000!}{1000!1000!}
\\ & \ds 
\frac{
\Omega_{[990, 1010]}
}{
\Omega_{[1000, 1000]}
}
= 
\frac{1000!1000!}{990!1010!}
=
\frac{991}{1001}\cdot \frac{992}{1002}\cdot \ldots \frac{1000}{1010},
\end{eqnarray*}
The ratio above satisfies 
\begin{equation*}
\left(\frac{99}{100}\right)^{10}
< 
\frac{
\Omega_{[990, 1010]}
}{
\Omega_{[1000, 1000]}
}
< 
\left(\frac{100}{101}\right)^{10}
\quad \therefore 
\frac{
\Omega_{[990, 1010]}
}{
\Omega_{[1000, 1000]}
}
\approx 
\left(\frac{99}{100}\right)^{10}
\end{equation*}
\item 
$N_\mathrm{total}=2\times 10^{23}$:
\begin{eqnarray}
\nonumber
& \ds
\Omega_{[0.99\NA, 1.01\NA]}
=
\frac{2\NA!}{(0.99\NA)!(1.01\NA)!},
\quad
\Omega_{[\NA, \NA]}
=
\frac{2\NA!}{\NA!\NA!}
\\ & \ds
     \frac{\Omega_{[0.99\NA, 1.01\NA]}
     }{
     \Omega_{[\NA, \NA]}
     }
 =
\frac{\NA!\NA!}{(0.99\NA)!(1.01\NA)!}
\approx
\left(\frac{99}{100}\right)^{0.01\NA}
\label{eq:Omega_99N_101N}
\end{eqnarray}
\end{enumerate}
Equation~\eqref{eq:Omega_99N_101N} indicates that the probability
to see 2~\% 
density difference ($0.99\rho_{0}$ and $1.01\rho_{0}$)
between the left and right for ideal gas systems approaches 
to zero for $\NA \gg 1$. In addition, by inserting 
Eq.~\eqref{eq:s=klnw} into Eq.~\eqref{eq:Omega_99N_101N}, 
it follows for the entropy difference $\Delta S$ between
$S_{[\NA, \NA]}$
and 
$S_{[0.99\NA, 1.01\NA]}$
that 
\begin{equation}
\Delta S \equiv 
    S_{[\NA, \NA]} 
    - 
    S_{[0.99\NA, 1.01\NA]}
    \approx
    0.01 \NA \kB \ln \left(\frac{100}{99}\right) > 0.
\end{equation}
This indicates that the entropy $S$ is the largest for 
uniform density, and also that $S$ is proportional to 
the number of molecules.
\par\noindent
 \hrulefill 
\par
Now we go back to the comparison of the Helmholtz 
free energy in Fig.~\ref{fig:drop-homogen}.
The entropy $S$ is larger for a system with 
homogeneous density than for an inhomogeneous system,
meaning that the right system is advantageous 
from the entropy aspect. 
At low temperature, $U$ contribution in 
Eq.~\eqref{eq:def_helmholtz} is dominant and 
the system tries to decrease $U$ to minimize $F$,
whereas at high temperature, the entropy contribution 
$TS$ overcomes the potential minimization effect.
%
This balance between the potential energy and the 
entropy governs the basic mechanism of phase separation 
as well as the surface tension. Note that for ideal gas 
systems, the potential contribution is constant (zero 
for the potential in Fig.~\ref{fig:potential}), and 
the gas molecules tries to fill the system with a 
homogeneous density to maximize entropy $S$.
Note as well that if a solid surface exists near 
the droplet at a temperature for the liquid-vapor phase 
separation, and the droplet is thermodynamically 
more stable on the surface, a hemispherical droplet 
is formed on the solid surface, and the equilibrium 
shape, \ie the contact angle $\theta$ in Fig.~\ref{fig:Young}, 
is determined so that the total free energy of the 
system may become minimum depending on the solid-fluid 
interaction strength.
%
%
\par
\section{Surface tension}
\begin{figure}[b]
\sidecaption
\includegraphics[scale=.50]{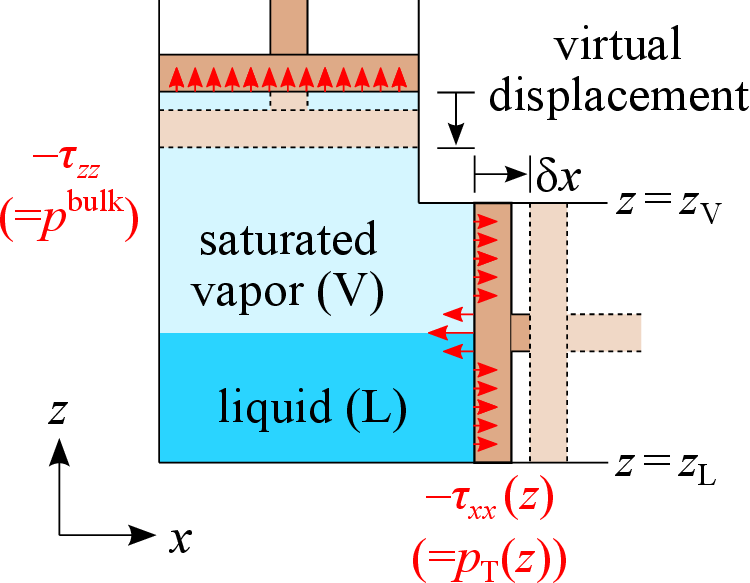}
%
%
\caption{Thought experiment of Bakker’s equation for a flat liquid-vapor interface. The red arrows denote the pressure, \ie normal fluid stress 
with its sign inverted, acting on the piston.
Reprinted with permission from Yamaguchi \textit{et al.}, J. Chem. Phys. 
150, 044701 (2019). Copyright 2019 Author(s), licensed under a Creative 
Commons Attribution (CC BY) license 4.0 License.}
\label{fig:bakker-LV}       
\end{figure}
\subsection{Bakker's equation}
As seen in the above example, we have to inevitably 
introduce the relation between the free energy of a 
fluid with interface and the local force to quantitatively 
evaluate the surface tension. For that purpose, 
we set a thought experiment with a flat liquid-vapor 
interface shown in Fig.~\ref{fig:bakker-LV}
for the basic connection between the thermodynamic 
work as energy and the surface tension.
Indeed, such flat interface can easily be achieved by MD 
simulations by using the periodic boundary condition 
in the surface lateral directions.
In this thought experiment, one piston is set normal to a flat
liquid-vapor interface, and it covers from 
$z = z_\mrl$ to $z = z_\mrv$ 
at the liquid and vapor bulk regions, respectively, across the plane
of the liquid-vapor interface. Another piston parallel to 
the interface is set in the vapor bulk far from the interface. 
Through simultaneous virtual infinitesimal displacements of 
the pistons, only the interface area can be changed without 
changing the liquid and vapor volumes, $V_\mrl$ and 
$V_\mrv$, respectively. Note that the change 
of bottom area in the figure is not considered.
Let $l$ be the depth 
normal to the $xz$-plane, and $\delta x$ be 
the corresponding displacement of the side piston, the 
change of interface area $A_\mrlv$ writes
\begin{equation}
    \delta A_\mrlv = l\delta x.
    \label{eq:deltaA}
\end{equation}
%
In order to let the internal energies before and 
after the displacement unchanged both for the liquid 
and vapor parts, this displacement must be done  
under constant temperature. Then, the net minimum 
mechanical work $\delta W$ exerted from the top 
and side pistons required for this change 
with quasi-static process can be associated 
with the change in the Helmholtz energy $F$.
Let $\glv$ be the LV interfacial energy per area, 
it follows for the quasi-static change that 
\begin{equation}
\delta F = \glv \delta A_\mrlv.
\end{equation}
Thus, $\glv$ writes
\begin{equation}
\glv = \left(\frac{\ptl F}{\ptl A_\mrlv}\right)_{N,V_\mrl,V_\mrv,T}.
\label{eq:glv=ptlF_ptlA}
\end{equation}
This means that surface tension is intrinsically
a thermodynamic force. 
\par
We further relate the mechanical stress changing as 
a continuous function with the LV interfacial tension $\glv$.
The fluid stress tensor is not isotropic at phase 
interfaces even at equilibrium as shown with the 
red arrows. Due to the one-dimensional feature of the 
system along the $z$-direction, the symmetric stress tensor $\bm{\tau}$ writes
\begin{equation}
    \bm{\tau} = 
    \left(\begin{matrix}
    \tau_{xx} & 0 & 0 \\
    0 & \tau_{yy} & 0 \\
    0 & 0 & \tau_{zz}
    \end{matrix}\right),
\end{equation}
where the surface-normal components $\tau_{zz}$ is 
constant in the entire region because of the force 
balance in the $z$-direction to be satisfy 
in the static equilibrium system for the present 
$z$-normal flat LV interface system.
This constant value 
is equal to the saturated vapor pressure $p^\mrb$
with its sign inverted, \ie
\begin{equation}
    \tau_{zz} = -p^\mrb.
    \label{eq:tauzz}
\end{equation}
On the other hand,
the surface-lateral diagonal components 
$\tau_{xx}$ and $\tau_{yy}$ satisfy
\begin{equation}
    \tau_{xx}(z) = \tau_{yy}(z) \equiv -p_\mathrm{T}(z),
    \label{eq:tauxxyy}
\end{equation}
where the surface tangential pressure is 
denoted by $p_\mathrm{T}(z)$, which is a unique 
function of position $z$ and
is equal to the isotropic vapor pressure 
$p^\mrb$ in the liquid and vapor bulks. 
We set the liquid and vapor 
bulk positions at $z=z_\mrl$ and $z=z_\mrv$,
respectively. Using these values, the changes 
of the volume and the Helmholtz free energy 
as the work upon the quasi-static displacement
$\delta x$ of the side piston are expressed by
\begin{equation}
\delta V = l\delta x \int_{z_\mrl}^{z_\mrv} dz,
\label{eq:deltaV}
\end{equation}
and
\begin{equation}
	dF\equiv \delta  W
	=
	p^\mrb \delta  V
	+ 
	l\delta x
	\int_{z_\mrl}^{z_\mrv} p_\mathrm{T}
	\left(z\right)dz,
	\label{eq:dF}
\end{equation}
respectively. By inserting Eqs.~\eqref{eq:deltaA} and 
\eqref{eq:tauzz}-\eqref{eq:dF} into Eq.~\eqref{eq:glv=ptlF_ptlA},
%
%
\begin{eqnarray}
	\nonumber 
	\glv &=& \ds
	\lim_{\delta x\rightarrow 0}\left.
	\frac{d F}{d  A_\mrlv}\right|_{N,V_\mrl,V_\mrv,T}
	\\ \ds \nonumber
	&=&
	p^\mrb\int_{z_\mrl}^{z_\mrv}
	dz
	-
	\int_{z_\mrl}^{z_\mrv}
	p_\mathrm{T}\left(z\right)dz
	=
	\int_{z_\mrl}^{z_\mrv}\left[
	p^\mrb - p_{T}\left(z\right)\right]
	dz
	\\ \ds
	&=&
	\int_{z_\mrl}^{z_\mrv}\left[
	\frac{\tau_{xx}(z)+\tau_{yy}(z)}{2} - \tau_{zz}\right]
	dz
	\label{eq:Bakker}
\end{eqnarray}
is derived. Equation~\eqref{eq:Bakker} is called Bakker's equation, 
which serves as the basic connection between the microscopic anisotropic 
stress distribution and the surface tension. Note that the 
integration range in Eq.~\eqref{eq:Bakker} is sometimes denoted by 
$\int_{-\infty}^{\infty}$;
however, the necessary condition for this range is that 
$z_\mrl$ and $z_\mrv$ cover the entire range of the 
interface with anisotropic fluid stress, \ie between 
the liquid and vapor bulks. 
\subsection{Molecular dynamics simulations}
\begin{figure}[b]
\includegraphics[scale=.60]{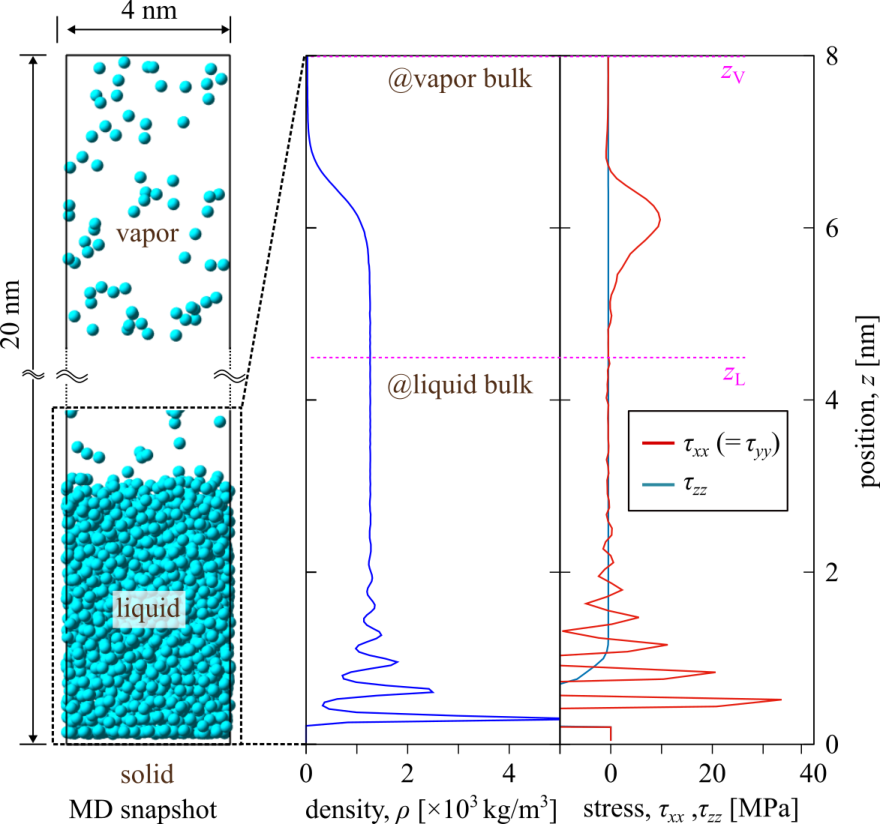}
\caption{
(Left) snapshot of an equilibrium molecular dynamics 
simulation system of Lennard-Jones molecules (argon) 
with flat solid-liquid and liquid-vapor interfaces.
(Middle) corresponding density distribution. 
(Right) distributions of stress diagonal components 
$\tau_{xx}$ and $\tau_{zz}$.
}
\label{fig:stress_dens_1d}       
\end{figure}
As mentioned in Section~\ref{sec:history}, 
molecular dynamics analysis is a powerful choice to 
overcome the theoretical difficulty of obtaining the 
equilibrium density distribution in liquid-vapor 
coexistence systems and resulting stress distribution 
to calculate the surface tension by Bakker's equation.
The left panel of Fig.~\eqref{fig:stress_dens_1d} 
shows a MD simulation system of Lennard-Jones (LJ) 
molecules (argon: 
$\sigma = 0.34$~nm, $\varepsilon=1.65\times 10^{-21}$~J in 
Eq.~\eqref{eq:LJ} and $m=39.948$~g/mol) 
with flat solid-liquid (SL) and liquid-vapor (LV) interfaces,
where 2000 argon molecules were confined in a rectangular 
simulation cell of $4\times 4 \times 20$~nm$^{3}$.
The periodic boundary conditions were imposed in the 
surface lateral $x$- and $y$-directions and the 
mirror boundary condition was set on the top boundary.
In addition, a solid wall modelled through an integrated
LJ potential was set on the bottom of the 
system to let the LJ-liquid be adhered on the solid.
The system was equilibrated at a temperature of 
$T=100$~K ($\kB T/\varepsilon=0.835$) 
until the apparent feature of the system 
did not change. With such setting, one can easily 
realize a quasi-one-dimensional equilibrium MD 
system with flat LV and SL interfaces normal 
to the $z$-direction.
\par
\begin{figure}[b]
\includegraphics[scale=.60]{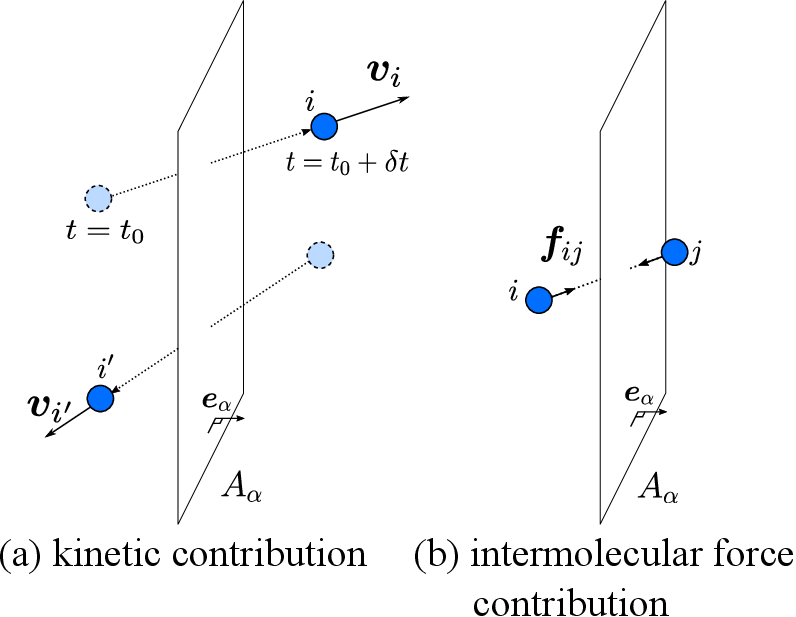}
\caption{
Schematics of the stress tensor calculation
based on the method of plane (MoP).
}
\label{fig:mop}
\end{figure}
It is easy to calculate the density of this equilibrium
system by taking the time-average of the mass in each
flat bin volume set normal to the $z$-direction with 
a small thickness of $\Delta z$.
On the other hand,  to calculate the stress, a 
method called the Method of Plane (MoP) is used,
where one sets control surfaces in the system 
instead of the bin volumes~\cite{Thompson1984, Yaguchi2010}. 
The local fluid stress tensor $\bm{\tau}(x,z)$ is  
calculated by setting $z$-normal and $x$- or $y$-normal 
flat bin faces in the system in the Cartesian coordinate. 
Figure~\ref{fig:mop} shows the schematics of 
the stress tensor calculation by the MoP.
The fluid stress tensor component $\tau_{\alpha\beta}$,
which expresses the stress in $\beta$-direction exerted on a 
surface element with an outward normal in $\alpha$-direction, 
is given by kinetic term $\tau_{\alpha\beta}^\mathrm{kin}$ 
and inter-molecular interaction term 
$\tau_{\alpha\beta}^\mathrm{int}$ 
as
\begin{equation}
\label{eq:stress_sum}
\tau_{\alpha\beta} = 
\tau_{\alpha\beta}^\mathrm{kin} + 
\tau_{\alpha\beta}^\mathrm{int}.
\end{equation}
In the MoP, the kinetic term on an $\alpha$-normal 
bin face with an area $A_\alpha$ is calculated by 
\begin{equation}
\label{eq:stress_kin}
\tau_{\alpha\beta}^\mathrm{kin} =
-\frac{1}{A_\alpha} 
\angb{
  \sum_{i \in \mathrm{fluid},\delta t}^{\mathrm{across }A_\alpha}
  \frac{
    \left(2\Theta(\bm{v}_{i}\cdot\bm{e}_\alpha)-1\right) 
    m_{i}\bm{v}_{i} \cdot \bm{e}_\beta
  }{\delta t}
},
\end{equation}
where $m_i$ and $\bm{v}_i$ denotes the mass and velocity 
vector of $i$-th fluid molecule, and $\bm{e}_\alpha$ and 
$\bm{e}_\beta$ are the unit normal vectors in $\alpha$- and 
$\beta$-directions, respectively. The angle brackets means 
the time average, and the summation 
$\sum_{i,\delta t}^{\mathrm{across }A_\alpha}$ 
is 
taken for every fluid molecule $i$ passing through 
the bin face within a time
interval of $\delta t$, which is equal to the time increment
for the numerical integration.
A switching function
$2\Theta(\bm{v}_{i}\cdot\bm{e}_\alpha)-1$, 
which gives $\pm 1$ depending on the sign 
of $\bm{v}_{i}\cdot\bm{e}_\alpha$ implemented 
through the Heaviside step function $\Theta$, 
is included in the RHS of Eq.~\eqref{eq:stress_kin}.
The meaning of the kinetic stress tensor in Eq.~\eqref{eq:stress_kin}
is shown with an example case with $\alpha=x$ and $\beta=y$.
Suppose that molecule $i$ with its $x$-directional velocity 
$v^{x}_{i}$ passes through a $x$-normal plane from left to 
right within a time interval $\delta t$ as in the 
upper molecule in Fig.~\ref{fig:mop}(a). This is only 
achieved for positive $v^{x(=\alpha)}_{i}$ satisfying 
$\bm{v}_{i}\cdot\bm{e}_{\alpha}>0$ with 
$\bm{e}_{\alpha}=(1,0,0)$, which corresponds to 
$2\Theta(\bm{v}_{i}\cdot\bm{e}_{\alpha})-1=1$.
Then, the $y$-directional momentum of the fluid
in the left is reduced by $m_{i}v^{y}_{i}$, \textit{i.e.}, 
is increased by $-m_{i}\bm{v}_{i}\cdot\bm{e}_{\beta}$
with $\bm{e}_{\beta}=(0,1,0)$. 
This change of momentum is equivalent to the impulse 
$\tau_{xy} A \delta t$ with the stress tensor. If the 
passage is in the opposite direction from right to left
 as in the lower molecule in Fig.~\ref{fig:mop}(a), 
then, $2\Theta(\bm{v}_{i}\cdot\bm{e}_{\alpha})-1=-1$ and 
a momentum of $m_{i}\bm{v}_{i}\cdot\bm{e}_{\beta}$ is 
given to the fluid in the left. From this formulation,
it is known that the kinetic contribution to surface 
normal stress $\tau_{\alpha \alpha}$ is always negative, \ie results in positive pressure, 
and this indeed corresponds to the ideal gas pressure $p$
given in Eq.~\eqref{eq:p_rho_msv}, which is proportional 
to the density $\rho$  on the face, and is also proportional 
to the temperature $T$ from Eq.~\eqref{eq:defT_micro}.
\par
On the other hand, the intermolecular interaction term in
Eq.~\eqref{eq:stress_sum} is given by
\begin{equation}
\label{eq:stress_int}
\tau_{\alpha\beta}^\mathrm{int} =
-\frac{1}{A_{\alpha}} 
\angb{
\sum_{(i,j)\in \mathrm{fluid}}^{\mathrm{across }A_\alpha}
\left(2\Theta(\bm{r}_{ij}\cdot\bm{e}_\alpha)-1\right) 
\bm{f}_{ij}\cdot\bm{e}_\beta
}, 
\end{equation}
where $\bm{r}_{ij}$ and $\bm{f}_{ij}$ denote the relative 
position vector $\bm{r}_{j} - \bm{r}_{i}$ and force vector 
exerted on molecule $j$ at position $\bm{r}_{j}$ from 
molecule $i$ at $\bm{r}_{i}$, respectively, and the summation 
$\sum_{(i,j)\in \mathrm{fluid}}^{\mathrm{across }A_\alpha}$
was taken for all line segments 
between $\bm{r}_{i}$ and $\bm{r}_{j}$ which crossed the bin face. 
In contrast to the kinetic contribution, this interaction 
contribution may give positive surface normal stress 
$\tau_{\alpha\alpha}$ on the bin face with an attractive 
interaction $\bm{f}_{ij}$.
\par
The middle and right panels of  Fig.~\ref{fig:stress_dens_1d} shows the density and surface-normal stress distributions, respectively.
As expected, the LV interface with a large change in the 
density has a certain thickness, and in this region with 
non-constant density, the normal stress $\tau_{xx}$ 
tangential to the interface is different from 
the bulk stress $\tau_{zz}$, which is constant 
in the whole system except near the solid at 
which the fluid is subject to an external force
from the solid.
Note that a remarkable wiggling structures in
the density and stress at bottom is specific to 
the solid-liquid interface, and this indeed is 
an interesting topic especially related to the wetting;
however, the effect vanishes as leaving away from 
the solid surface, and we mainly focus on the liquid-vapor 
interface. 
It is really amazing that the frontiers could expect 
the stress distribution at least around the LV 
interface without MD simulations. Also note 
that even by such great scientists, quantitatively 
evaluating the density and stress distributions 
only from the intermolecular potential was 
still difficult, and that we have a powerful tool 
of MD simulations to complement the missing piece.
Related to this point, 
calculating the SL (and SV) interfacial tension
from the stress distribution obtained through 
the MD simulations toward the understanding of
wetting is a hot topic at 
present~\cite{Surblys2014,Yamaguchi2019,Kusudo2021}; 
however,  I will not describe about this in detail
considering the scope of this article.
\par
\begin{figure}[b]
\includegraphics[scale=.75]{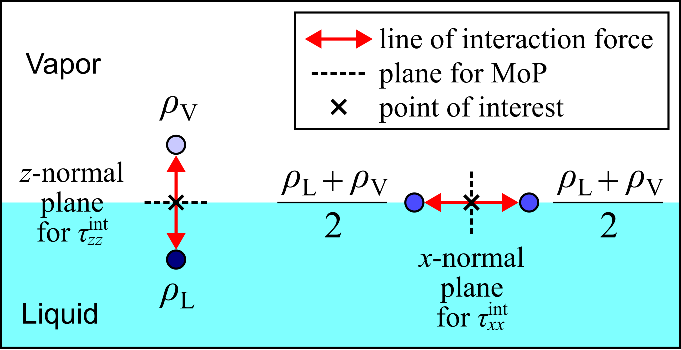}
\caption{
The simplest model to explain the stress anisotropy at the interface. The interaction terms $\tau^\mathrm{int}_{zz}$ and $\tau^\mathrm{int}_{xx}$ on the point of interest (black crosses) are represented by the interaction between two points 
having densities $(\rhoL, \rhoV)$, and $(\frac{\rhoL+\rhoV}{2},\frac{\rhoL+\rhoV}{2})$, respectively connected by the 
double-headed red arrows.
}
\label{fig:stress-aniso-model}
\end{figure}
A very simple and intuitive model to explain the stress anisotropy at the liquid-vapor interface, \ie $\tau_{zz}<\tau_{xx}$, is given here. 
Suppose $z$-normal and $x$-normal planes on a $z$-normal flat LV interface for the calculation of the normal stress components $\tau_{zz}$ and $\tau_{xx}$, respectively based on the MoP as shown in Fig.~\ref{fig:stress-aniso-model}. In this model, we assume average densities $\rhoL$ and $\rhoV (< \rhoL)$ and 
$\frac{\rhoL+\rhoV}{2}$ in the liquid and vapor bulks and on the interface, respectively for simplicity. In addition, we introduce a mean-field like approach to the MoP here; instead of counting the molecules passing through the MoP plane for the kinetic stress term $\tau^\mathrm{kin}$ in Eq.~\eqref{eq:stress_kin} or summing up the intermolecular interaction force for the line segment between two molecules crossing the MoP plane for the interaction stress term 
$\tau_\mathrm{int}$ in Eq.~\eqref{eq:stress_int}, we evaluate $\tau^\mathrm{kin}$ with the density on the point of interest (black cross) on the MoP plane, 
and $\tau^\mathrm{int}$ with the two densities on the points 
which sandwich the point of interest (double-headed red arrow)
in Fig.~\ref{fig:stress-aniso-model}. As easily imagined, to correctly calculate the latter interaction term of $\tau^\mathrm{int}$, complicated space 
integration for the two points using a position- and direction-dependent radial distribution function is needed, 
and this is indeed the difficulties 
that the great scientists could not solve analytically, but here 
we try to represent the interaction term by the most probable single distance in the present simplest model.
The kinetic terms $\bm{\tau}^\mathrm{kin}$ is isotropic in equilibrium 
systems and is given by
\begin{equation}
\label{eq:stress_kin_rho}
	\bm{\tau}^\mathrm{kin} 
 = -\frac{N \kB T}{V}\bm{I} = -\rho \frac{\kB T}{m}\bm{I},
\end{equation}
where $\bm{I}$ denotes the identity tensor.
Equation~\eqref{eq:stress_kin_rho} corresponds to the 
ideal-gas pressure, which can be interpreted with 
Eqs.~\eqref{eq:defT_micro} and 
\eqref{eq:stress_kin} as follows: 
fast molecules giving a momentum of 
$m\bm{v}_{i}$ in Eq.~\eqref{eq:stress_kin} pass through the plane 
more with proportional to $|\bm{v}_{i}|$. 
From Eq.~\eqref{eq:stress_kin_rho},
$\tau^\mathrm{kin}_{zz}$ and $\tau^\mathrm{kin}_{xx}$ 
are identical and those at the LV interface at $z=z_\mathrm{LV}$ are expressed by
\begin{equation}
\tau^\mathrm{kin}_{zz}(z_\mathrm{LV})
 = 
\tau^\mathrm{kin}_{xx}(z_\mathrm{LV})
 = 
 -\frac{\rhoL + \rhoV}{2} \frac{\kB T}{m}.
\end{equation}
On the other hand, we assume that the interaction 
term is represented by a single line of interaction force of 
the most probable distance (double-headed red arrows) which gives 
an intermolecular force $f_\mathrm{rep}(z)$, then the interaction term $\tau^\mathrm{int}_{\alpha\alpha}$ writes
\begin{equation}
	\tau^\mathrm{int}_{\alpha\alpha}(z) \equiv 
	-\rho_{1}\rho_{2} c(z)f_\mathrm{rep}(z),
 \label{eq:tauint_rep}
\end{equation}
where $\rho_{1}$ and $\rho_{2}$ are the density of the two points
sandwiching the point of interest, and $c (>0)$ is a certain positive 
coefficient for the model due to this representation.
Note that negative $f_\mathrm{rep}(z)$ corresponds to 
attractive force across the MoP plane to give positive normal stress.
Since $\tau_{zz}$ must satisfy Eq.~\eqref{eq:tauzz}, 
$\tau_{zz}(z_\mathrm{LV})$ at the interface $z=z_\mathrm{LV}$ given by 
\begin{equation}
\tau_{zz}(z_\mathrm{LV})
=
\tau_{zz}^\mathrm{kin}(z_\mathrm{LV})
+
\tau_{zz}^\mathrm{int}(z_\mathrm{LV})
=
- \frac{\rhoL + \rhoV}{2} \frac{\kB T}{m}
-
\rhoV\rhoL c(z_\mathrm{LV})f_\mathrm{rep} (z_\mathrm{LV})
\label{eq:tauzz_zLV}
\end{equation}
with $(\rho_1,\rho_2)= (\rhoL,\rhoV)$ in Eq.~\eqref{eq:tauint_rep}, is equal to 
$\tau_{zz}(z_\mathrm{V})$ in the vapor bulk at $z=z_\mathrm{V}$,
corresponding to the saturated vapor pressure. This is given by 
\begin{equation}
\tau_{zz}(z_\mathrm{V})
=
\tau_{zz}^\mathrm{kin}(z_\mathrm{V})
+
\tau_{zz}^\mathrm{int}(z_\mathrm{V})
=
 -{\rhoV} \frac{\kB T}{m}
-
\rhoV^{2} c(z_\mathrm{V}) f_\mathrm{rep} (z_\mathrm{V})
\approx
-{\rhoV} \frac{\kB T}{m},
\label{eq:tauzz_vbulk}
\end{equation}
where the interaction term is assumed to be zero 
for the last approximation
considering that the intermolecular interaction in the vapor is negligible. 
Thus, it follows 
\begin{equation}
  c(z_\mathrm{LV})
  f_\mathrm{rep}(z_\mathrm{LV})
  =
  -\frac{\rhoL-\rhoV}{2}\frac{\kB T}{m}
  < 0,
\end{equation}
indicating that $f_\mathrm{rep}(z_\mathrm{LV})$ is negative, \ie the representative interaction across 
the MoP plane at the LV interface is attractive.
Assuming as well that $c(z_\mathrm{LV})f_\mathrm{rep}(z_\mathrm{LV})$ is independent of the direction at $z_\mathrm{LV}$, then the normal stress lateral to the interface writes 
\begin{equation}
\left.\tau_{xx}\right|_{z=z_\mathrm{LV}}
=
- \frac{\rhoL + \rhoV}{2} \frac{\kB T}{m}
-
\left(\frac{\rhoV+\rhoL}{2}\right)^{2} c(z_\mathrm{LV})f_\mathrm{rep} (z_\mathrm{LV}).
\label{eq:tauxx_zLV}
\end{equation}
Using the relation between the arithmetic and geometric means
\begin{equation}
\rhoL \rhoV
<
\left(
\frac{\rho_\mathrm{L} + \rho_\mathrm{V}}{2}
\right)^{2},
\end{equation}
it follows for $\tau_{zz}(z_\mathrm{LV})$ and $\tau_{xx}(z_\mathrm{LV})$ in Eqs.~\eqref{eq:tauzz_zLV} and \eqref{eq:tauxx_zLV}, respectively that 
\begin{equation}
    \tau_{zz}(z_\mathrm{LV})<\tau_{xx}(z_\mathrm{LV}).
\end{equation}
This inequality is a simple model to explain the stress 
anisotropy shown in Fig.~\ref{fig:stress_dens_1d}, 
due to the density anisotropy, and this is the fundamental 
basis of Bakker's equation~\eqref{eq:Bakker}.
\par
\begin{figure}[b]
\includegraphics[scale=.45]{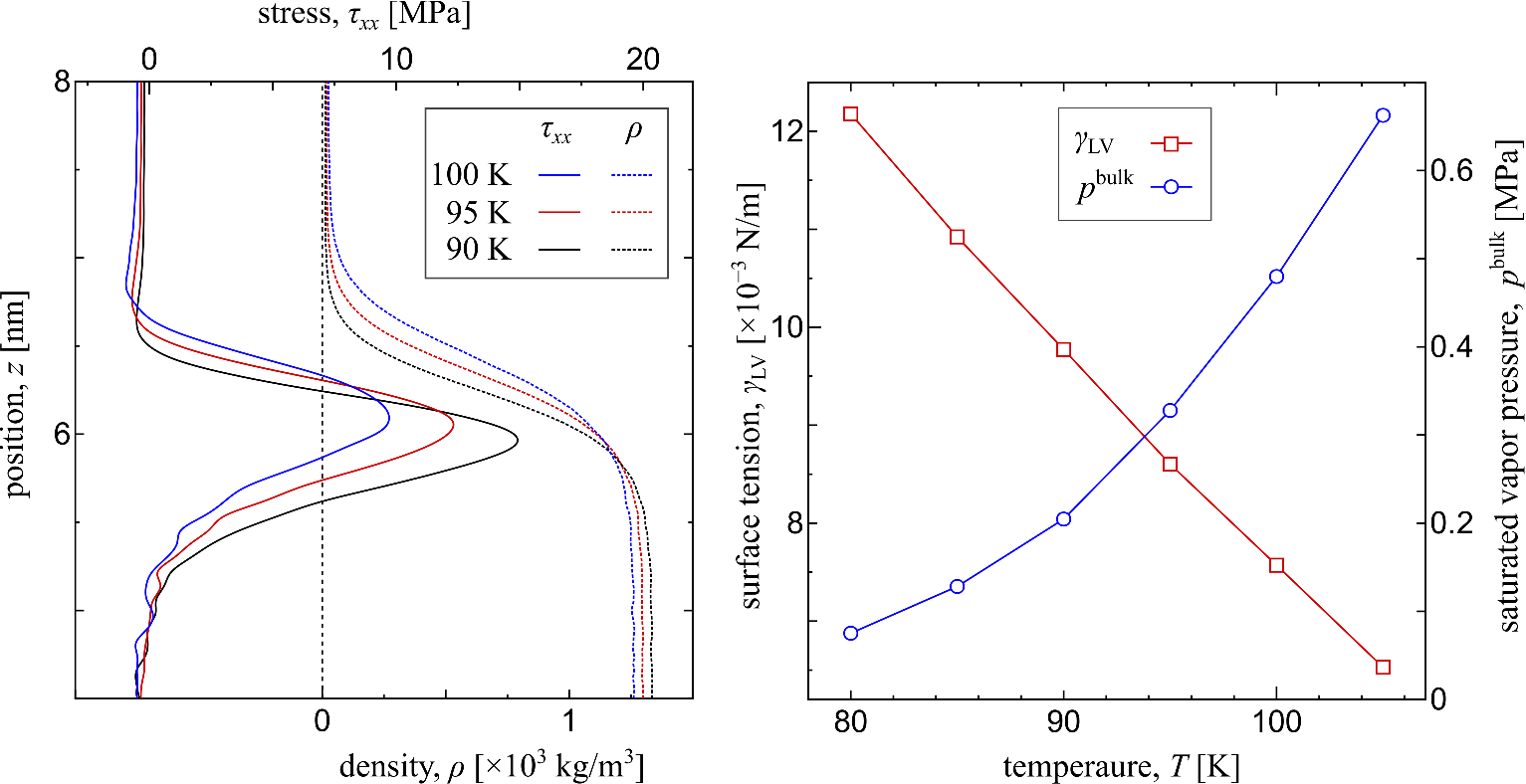}
\caption{
(Left) distributions of the normal stress $\tau_{xx}$ 
tangential to the interface and density $\rho$ 
at three different temperature $T=90, 95$ and 100~K 
enlarged around the LV interface.
(Right) liquid-vapor interfacial tension $\glv$ and 
$p^\mrb$ at different temperature $T$.
}
\label{fig:temp-glv-pblk}
\end{figure}
The left panel of Fig.~\eqref{fig:temp-glv-pblk} shows the 
distributions of the normal stress $\tau_{xx}$ 
tangential to the interface and density $\rho$ 
at three different temperature $T=90, 95$ and 100~K
enlarged around the LV interface. As the temperature 
increases, the stress difference between the values of 
$\tau_{xx}(z)$ at the interface and away from the interface  
becomes smaller. The latter away from the interface is equal to 
and $\tau_{zz}$, which is equal to the constant stress 
value away from the LV interface as shown in Fig.~\ref{fig:stress_dens_1d}.
From Bakker's equation~\eqref{eq:Bakker}, 
this means that the LV interfacial tension $\glv$ decreases 
with the temperature rise. Note also that the bulk liquid density 
decreases with the the temperature rise, and this is related 
to the increase of the average distance upon temperature rise 
indicated in Fig.~\ref{fig:phasechange}. In addition, the bulk
stress decreases with the temperature rise, and this means 
that the saturated vapor pressure $p^\mrb_\mrv$ given 
by Eq.~\eqref{eq:tauzz} increases with temperature. 
Such basic features are indeed realized in the MD systems.
\par
The right panel of Fig.~\ref{fig:temp-glv-pblk}
shows the LV interfacial tension $\glv$ and $p^\mrb$ 
at different temperature $T$. As expected from the right 
panel, $\glv$ decreases and $p^\mrb$ increases with the 
temperature increase. The vapor pressure is above 
the atmospheric pressure, and the extraordinary high
anisotropic local stress $\tau_{xx}$ gives the interfacial 
tension around 10~mN/m, and the resulting temperature 
dependence of about $-0.2$~mN/m$\cdot$ K. Of course, the absolute 
value and the coefficient of the temperature dependence depend on the 
molecular type and temperature range, but in general, 
this non-negligible temperature gradient gives the 
local stress gradient around the interface known as 
the temperature Marangoni effect, which drives the 
liquid flow from a higher-temperature region to 
a lower-temperature region.
\section{Concluding remarks}
Surface is physically as well as theoretically at the boundary of macroscopic and microscopic physics, understanding of surface tension 
intrinsically needs multiple knowledge of thermodynamics, statistical mechanics as well as continuum mechanics. In this article, after introducing the interpretation of the phase change, liquid-vapor coexistence and interfacial tension from these points of view, molecular dynamics results were shown which indicate the existence of microscopic stress anisotropy at the interface as the origin of the macroscopic stress surface tension tangential to the interface. In addition, a simple model to explain the stress anisotropy due to the density anisotropy was provided. 
At present, molecular dynamics as a powerful tool is available, and I hope that unsolved issues about this fascinating physics, especially related to the Marangoni effect and wetting, will be elucidated.
\begin{acknowledgement}
The simulation results in this article are provided 
by Minori Shintaku and Yule Ding, former members of the author's group. I sincerely appreciate their cooperation.
\end{acknowledgement}
\end{document}